\documentclass[aps,prl,preprint,groupedaddress]{revtex4}

\usepackage{graphicx}

\begin{document}


\title{Anisotropy of superconductivity of as-grown MgB$_2$ thin films by molecular beam epitaxy}



\author{Y. Harada}
 \affiliation{Advanced Science and Technology Institute of Iwate, Iwate Industrial Promotion Center, \\
Iioka shinden 3-35-2, Morioka, 020-0852, Japan}
\author{M. Udsuka, Y. Nakanishi, and M. Yoshizawa}
 \affiliation{Graduate School of Frontier Materials and Functional Engineering, 
        Iwate University, \\ 
        Ueda 4-3-5, Morioka, 020-8551, Japan}



\date{\today}

\begin{abstract}
Superconducting thin films of magnesium diboride (MgB$_2$) were prepared on MgO (001) substrate by a molecular beam epitaxy (MBE) method with the co-evaporation conditions of low deposition rate in ultra-high vacuum. 
The structural and physical properties of the films were studied by RHEED, XRD, XPS, resistivity and magnetization measurements.
All films demonstrated superconductivity without use of any post-annealing process.
The highest {\it T}$_{c,onset}$ determined by resistivity measurement was about 33K in the present samples.
Anisotropic superconducting properties were evaluated by the resistivity and magnetic measurements.
We will discuss the anisotropy of superconductivity for as-grown MgB$_2$ thin films.
\end{abstract}


\maketitle


\section{Introduction}

The discovery of superconductivity at 39K in magnesium diboride (MgB$_2$) has attracted great interest in science and technology since it shows the highest transition temperature ({\it T}$_c$) among intermetallic compounds \cite{aki1}. 
There have already been several reports on the preparation technique of MgB$_2$ thin film so far \cite{kang1, Eom1}.
Many of them require a post-annealing process to improve their physical properties and attain superconductivity.
As-grown process without the use of any post-annealing process is strongly desired for the fabrication of tunnelling junctions and multilayers. 
Several groups have reported as-grown superconducting MgB$_2$ thin films so far \cite{Ueda1, Jo1, Erven1}.
In as-grown method, low temperature synthesis, is needed to depress high volatility of Mg, makes difficult to produce MgB$_2$ thin film with high crystallinity.
It is considered that MgB$_2$ needs to be grown up in vacuum as high as possible to get good crystallinity, because the quality of MgB$_2$ thin film is considerably affected by residual gases \cite{ueda2}.
This method is a promising way for device application.
Therefore, it is desired to establish the low rate deposition technique at low growth temperature to produce high quality films.

The evaluation of the anisotropy in superconducting materials is very important not only for the basic understanding of this material but also for its potential applications because anisotropy strongly affects the flux pinning, critical currents, and electronic device limit. 
Since MgB$_2$ consists of alternating hexagonal layers of Mg atoms and graphite-like honeycomb layers of B atoms, electronic anisotropy has been predicted by theoretical calculations \cite{An1,kortus1}. 
However, only few indirect measurements have been made to determine the anisotropy, e.g., on single crystal \cite{Xu1}, and c-axis oriented thin films of MgB$_2$ \cite{Pat1}. 
The anisotropy ratio, $\gamma=${\it H}$_{C2,c}$(0)/{\it H}$_{C2,ab}$(0), were reported as 2.6 and 1.8-2.0, respectively. 

We reported previously the synthesis of thin film by molecular beam epitaxy (MBE) in the co-evaporation conditions of low growth temperature, low deposition rates and ultra high vacuum \cite{harada1}.  
In this paper, we present the optimizing of the co-evaporation conditions for as-grown superconducting thin film of MgB$_2$ using co-evaporation method. 
And anisotropic superconducting properties were evaluated by the resistivity and magnetic measurements.

\section{Experimental}

The MgB$_2$ films were grown on MgO(001) substrates in an MBE chamber with a base pressure of 4$\times 10$$^ {-10}$ Torr.
Pure metal of Mg and B were used as the evaporation sources. 
The both purities of Mg and B were 3N.
Mg was evaporated from a Knudsen cell, and B by electron-beam.
The deposition rate was controlled by a quartz-crystal monitor (QCM) with the flix ratio of Mg to B varied from 9 to 10 times as high as the nominal flux ratio to compensate Mg loss.
The deposition rate of B was varied from 0.5\AA/sec to 0.3\AA/sec.
The thickness of films was typically 1000\AA.
The growth temperature ({\it T}$_S$) was varied from 300$^\circ$C to 200$^\circ$C. 
The background pressure during the growth was better than 4$\times 10$$^{-10}$Torr.
All of the films were covered by 50\AA-thick Mg cap layer to avoid oxidation.
The crystal structure was characterized by in-situ reflection high-energy electron diffraction (RHEED) and ex-situ X-ray diffraction (XRD: 2$\theta-\theta$ scan).
The composition of the films was investigated by X-ray Photoelectron Spectroscopy (XPS).
The resistivity and magnetic susceptibility measurements were carried out by the standard four-probe technique and SQUID magnetometer, respectively.

\section{Results and Discussion}

We examined the characteristics of a thin film made under a range of deposition conditions on a MgO(001) substrate.
Firstly, we examined MgB$_2$ thin films made by using a relatively low deposition temperature of 100-300$^\circ$C.
The deposition rate of B was fixed at 0.5\AA/sec.
Figure 1(a) shows the resistivity as a function of temperature curve of the MgB$_2$ thin films grown at 200$^\circ$C and 250 $^\circ$C with an Mg/B ratio of 9-10.
The {\it T}$_{c,onset}$-{\it T}$_{c,offset}$ was 33.1-31.1K and 31.3-28.3K, respectively.
Here, {\it T}$_{c,onset}$ was defined as the temperature at resistance abruptly to drop, and the {\it T}$_{c,offset}$ was defined as the temperature at which resistivity reached zero.
The normal state resistivity at room temperature (RT) was 159.5 (250$^\circ$C) and 93.1 $\mu\Omega$cm(200$^\circ$C), respectively.
It should be marked that the resistivity of the films grown at 200$^\circ$C is lower than those of at 250$^\circ$C.
This result can be ascribed to the reduction of grain boundary, that is, enlargement of grain size.

The dc magnetic properties were measured with a superconducting quantum interference device magnetometer (MPMS-5, Quantum design) at an applied field perpendicular to the film plane.
The size of samples was 3$\times 3$mm$^2$.
Figure 1(b) shows the temperature dependence of the zero-field-cooled and field-cooled dc magnetization ({\it M}-{\it T}) curves of the MgB$_2$ thin films grown at 200$^\circ$C and 250$^\circ$ under a field of 1mT.
The {\it M}-{\it T} curves exhibit the {\it T}$_c$ around 26K (250$^\circ$C) and 31K (200$^\circ$C), where the growth temperature are designated in parenthesis.

Figure 2 shows the substrate temperature ({\it T}$_s$) dependence of the {\it T}$_c$.
Between 200$^\circ$C and 250$^\circ$C, the films showed superconducting transition above 30K,whereas that made at 300$^\circ$C showed no superconductivity.
This indicates that the growth temperature limit is 300$^\circ$C for the as-grown superconducting MgB$_2$ thin films by these co-evaporation conditions.
From these results, we have chosen the typical growth conditions as {\it T}$_s$=200$^\circ$C.

To establish optimal conditions, we studied the dependence of B deposition rate on the film properties.
In this series of films, the growth temperature was fixed at 200$^\circ$C.
The B deposition rate was decreased from 0.5 to 0.1\AA/sec to get lower vacuum conditions.
The Mg/B ratio was fixed at 9-10.

Figure 3(a) shows the resistivity as a function of temperature curve at various B deposition rates.
The {\it T}$_{c,onset}$-{\it T}$_{c,offset}$ was 31.3-28.1K (0.5\AA/sec), 32.6-31.9K (0.3\AA/sec) and 30.6-26.2(0.1\AA/sec), respectively.
The parenthesis mean the B deposition rates.
The transition width ($\Delta${\it T}$_c$) at 0.3\AA/sec had a very low value of 0.7K.
In 0.1\AA/sec, a clear step like transition were found.
This pattern suggests that the possibility of two different MgB$_2$ phase exists in this film.

Figure 3(b) shows the {\it M}-{\it T} curves, which B deposition rate was 0.3 and 0.5\AA/sec.
The {\it M}-{\it T} curves exhibit the {\it T}$_c$ around 30K.
As these results, the quality and superconducting properties of MgB$_2$ thin film made at 0.3\AA/sec were improved than other deposition rate films.
Therefore, we have chosen the typical growth conditions; B deposition rate: 0.3\AA/sec.
 
Next, we studied anisotropy of superconductivity of as-grown MgB$_2$ thin film.
New film was prepared at {\it T}$_s$=200, B deposition rate of 0.3\AA/sec and Mg deposition rate of around 3\AA/sec.
The structure and crystallinity were characterised by RHEED and XRD.
Figure 4(a) shows the XRD pattern of the MgB$_2$ thin film. 
XRD results show the (00l) diffraction peak of the MgB$_2$.
The RHEED pattern showed a ring pattern. 
The composition of the films was investigated by XPS. 
From the XPS depth profile shown in Figure 5, it was found that a MgB$_2$ thin film is slightly Mg deficient (i.e., 25.0\% Mg and 71.3\% B). 

The dc magnetic properties were studied with a superconducting quantum interference device magnetometer (MPMS-5. Quantum Design) at an applied field parallel ({\it H}$\|c$) or the perpendicular to the film plane ({\it H}$\|$ab). 
Figure 6(a) shows {\it M}-{\it T} curves of the MgB$_2$ thin film under a field of 1mT along the perpendicular to the film plane.
The {\it M}-{\it T} curves exhibit the same superconducting transition temperature of {\it T}$_C$ around 31K. 
The magnetization of the thin film, as a function of an applied field up to 3T ({\it M}-{\it H} curve), was measured at several temperatures.
Figure 6(b) shows the {\it M}-{\it H} curves at 5K for applied fields up to 3T at {\it H}$\|c$ and {\it H}$\|$ab.
They show the typical hysteresis loop, indicating the characteristic curve of type-I\hspace{-.1em}I superconductors. 
{\it M}-{\it H} curves exhibit a remarkable anisotropy between parallel and perpendicular to the film plane.

To estimate the critical current density({\it J}$_C$), we measured {\it M}-{\it H} curves of the same samples as a function of temperature.
The {\it J}$_C$ was calculated based on the Bean model \cite{gyorgy1}, {\it J}$_C$=30$\Delta mV^{-1}r^{-1}$, where $\Delta m$ is the width of the hysteresis loop, {\it V} is the film volume, and r is the sample half-width.
At zero field, the {\it J}$_{C,c}$ was estimated to be 7.1$\times10$$^7$A/cm$^2$ at 5K and $\sim$5.1$\times 10$$^7$A/cm$^2$ at 20K, respectively from the {\it M}-{\it H} curves of {\it H}$\|$c.
The {\it J}$_{C,ab}$ was $\sim$4.7$\times10$$^6$A/cm$^2$ at 5K and $\sim$3.2$\times 10$$^6$A/cm$^2$ at 20K, respectively from the {\it M}-{\it H} curves of {\it H}$\|$ab.
Here, the sample size was employed for the calculation of {\it J}$_C$, instead of the grain size.
 
Figure 7 shows the temperature dependence of the resistivity under the selected magnetic fields up to 14T for {\it H}$\|$c and {\it H}$\|$ab. 
The {\it T}$_{C,onset}$ is estimated to be 31.2K at 0T, which is consistent with the result of {\it M}-{\it T} measurements.
The transition width gradually becomes broader with increasing magnetic field from about 1K in 0T to 6.0K in 12T.
This broadening is smaller compared to previous reported one of MgB$_2$ wires \cite{canfield1}, bulk samples \cite{finnemore1}, or thin films \cite{Jo1}.

From the {\it M}-{\it H} curves, the lower critical fields, {\it H}$_{C1}$ for {\it H}$\|$c and {\it H}$\|$ab, were defined by the magnetic fields where the initial slope of {\it M}$_{up}$ curve meets the extrapolated curve of ({\it M}$_{up}$ + {\it M}$_{down}$)/2.
The temperature dependence of {\it H}$_{C1}$ is plotted as shown in Figure 8. 
Extrapolation of the plot to zero temperature gives the {\it H}$_{C1,c}$(0) values of 14mT.
This {\it H}$_{C1}$(0) gives the penetration depth $\lambda$ of 21.7nm at zero temperature based on Ginzburg-Landau (GL) formulas, $\mu${\it H}$_{C1}\sim\Phi _0$/$\pi\lambda^2$. 

The {\it H}$_{C2}$ for {\it H}$\|$c and {\it H}$\|$ab, were determined by the resistive onset temperature in the Figure 7.
The {\it H}$_{C2}$({\it T}) curves for {\it H}$\|$c and {\it H}$\|$ab show a linear behavior in the temperature region far from {\it T}$_C$.
These results give an approximate slope at {\it T}$_C$ of {\it dH}$_{C2}$/{\it dT}=0.61(perpendicular) and 1.1(parallel) T/K determined from the onset.
Therefore, a linear extrapolation to zero temperature gives the {\it H}$_{C2,c}$(0) and {\it H}$_{C2,ab}$(0) values of 19.6T and 34.5T, respectively.
Assuming the dirty limit of the type-I\hspace{-.1em}I superconductor, the {\it H}$_{C2}$(0) values \cite{werthamer} is expressed as {\it H}$_{C2}$(0)$\approx$0.691$|${\it dH}$_{C2}$/{\it dT}$|$$\times {\it T}$$_C$. 
Thus, {\it H}$_{C2,c}$(0) and {\it H}$_{C2,ab}$(0) are estimated to be 13.2T and 23.6T, respectively.
The present values of {\it H}$_{C2}$(0) in MgB$_2$ thin films are larger than previous ones, for example, grown by the two-step method \cite{Eom1} and reported as-grown method \cite{Jo1}.
Using the anisotropic GL formulas: {\it H}$_{C2,c}$=$\phi$/(2$\pi$$\xi$$_{ab}$\hspace{-.1em}$^2$) and {\it H}$_{C2,ab}$=$\phi$/(2$\pi$$\xi$$_{ab}$$\xi$$_c$), the GL coherence length $\xi$$_c$(0) and $\xi$$_{ab}$(0) at zero temperature were calculated to be 2.8nm and 5.0nm, respectively.
These values are larger than the typical values observed for high-temperature superconductors (HTS) and comparable to those estimated from the high-pressure sintered polycrystalline sample \cite{aki1} and single crystal \cite{Xu1}.
From the upper critical field {\it H}$_{C2}$, the anisotropy ratio [$\gamma$={\it H}$_{C2,ab}$(0)/{\it H}$_{C2,c}$(0)] was estimated to be 1.78, that is an anisotropy of the coherence length $\xi$$_{ab}$(0)/$\xi$$_c$(0)$\approx$1.78.
This value of $\gamma$ is comparable to the reported value of 1.8-2.0 for the c-axis oriented MgB$_2$ thin films by two-step method \cite{Pat1}.
The important physical values, {\it H}$_{C2}$, $\xi$, {\it J}$_C$, {\it H}$_{C1}$, $\lambda$ and $\gamma$ of our films are summerized in Table 1.  

In summary, we have grown the as-grown MgB$_2$ thin films on MgO(001) substrate by MBE in the conditions of low temperature, low deposition rate and ultra-high vacuum.
Currently, B deposition rate is 0.3\AA/sec and growth temperature is 200$^\circ$C are best conditions to make the high quality MgB$_2$ thin film.
The films exhibit clear (00l) peaks of MgB$_2$.
However, the composition is not yet good enough to get perfectly stoichiometric films.
A superconducting transition with the onset temperature around 32K was confirmed through the resistivity and magnetization measurements.
We evaluated upper critical field anisotropy ratio of as-grown MgB$_2$ thin films for the first time.
The upper critical field anisotropy ratio of as-grown MgB$_2$ thin film, $\gamma$={\it H}$_{C2,ab}$(0)/{\it H}$_{C2,c}$(0) of 1.78, was estimated from the magnetic field-temperature phase diagram for the first time.
The observed anisotropy is comparable to the values observed for MgB$_2$ thin films by two-step method and single crystal
These results demonstrated that this as-grown MgB$_2$ thin film using MBE would be promising for the device application.

\begin{acknowledgments}
The authors would like to thank Dr. A. Yamaguchi at Iwate University for helpful discussion.
We are grateful to M. Nakamura, T. Kumagai, and M. Oikawa for their help in the transport measurements and the operation of the cryogenic apparatus.
The measurements have been carried out in the Cryogenic Division of the Center for Instrumental Analysis, Iwate University.
\end{acknowledgments}

\bibliography{mgb2}

\newpage



%

\begin{table}[tp]
\begin{center}
\label{table1}\caption[super]{Superconducting properties of as-grown MgB$_2$ thin films in this work}
\smallskip
\begin{tabular}{cc}\hline
{\it H}$_{C2,c}$ & 19.6T\\
{\it H}$_{C2,ab}$ & 34.5T\\
{\it H}$_{C2,c}$(dirty limit) & 13.2T\\
{\it H}$_{C2,ab}$(dirty limit) & 23.6T\\
$\xi _c$(0) & 2.8nm\\
$\xi _{ab}$(0) & 5.0nm\\
{\it J}$_{C,c}$(5K) & 7.1$\times 10^7A/cm^2$\\
{\it J}$_{C,ab}$(5K) & 4.7$\times 10^6A/cm^2$\\
{\it H}$_{C1,c}$ & 14mT \\
$\lambda _c$ & 21.7nm\\\hline
\end{tabular}
\end{center}
\end{table}

\newpage

\begin{figure}[htp]
	\includegraphics[width=0.8\linewidth]{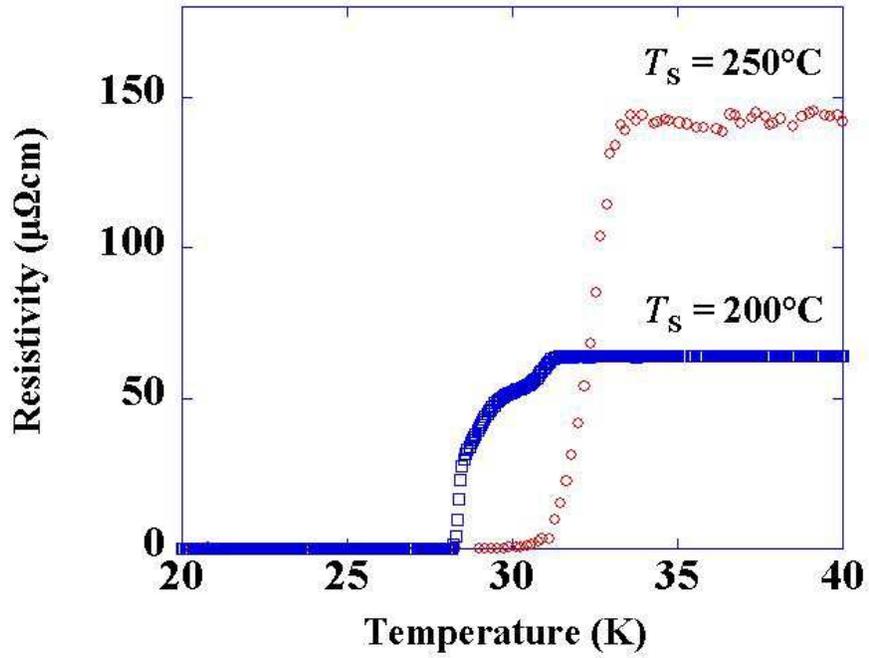}
	\includegraphics[width=0.8\linewidth]{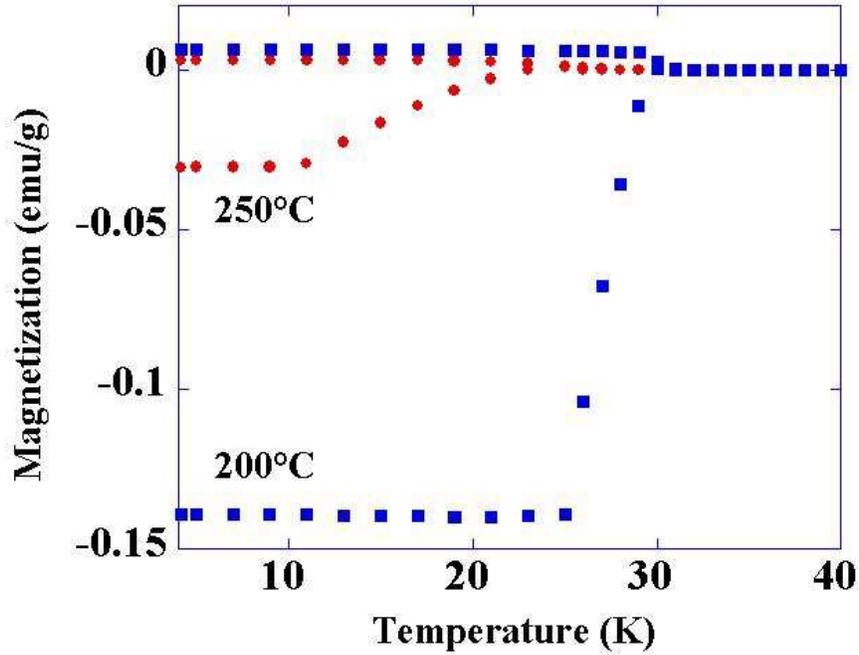}
	\caption{(a) Resistivity as a function of temperature of MgB$_2$ thin films grown at 250$^\circ$C and 200$^\circ$. (b) Magnetization of these films as a function of temperature after cooling under zero field and under a field of 1mT.}
\end{figure}
\begin{figure}[tp]
	\includegraphics[width=0.8\linewidth]{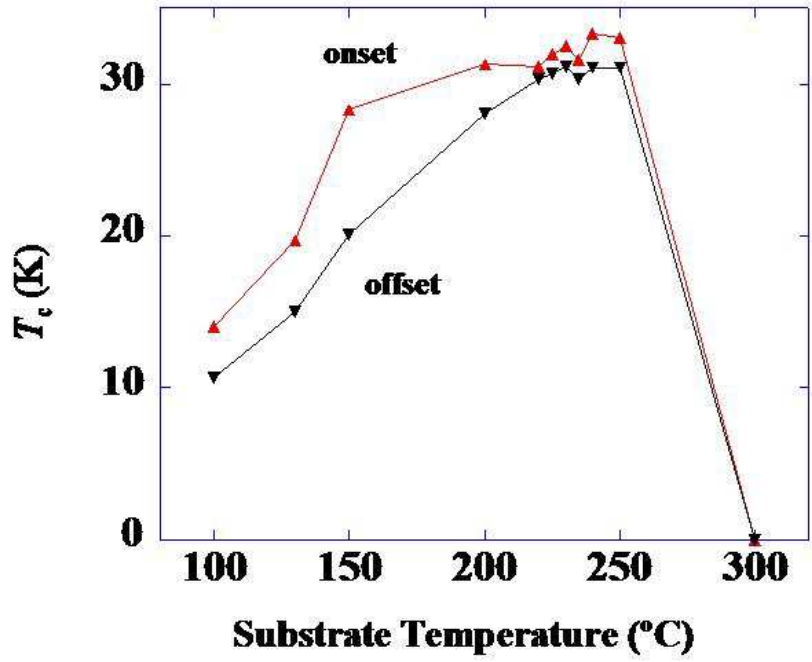}
	\caption{Substrate temperature dependence of critical temperature. B deposition rate: 0.5\AA/sec, Mg deposition rate: 4.3\AA/sec}
\end{figure}
\begin{figure}[tp]
	\includegraphics[width=0.8\linewidth]{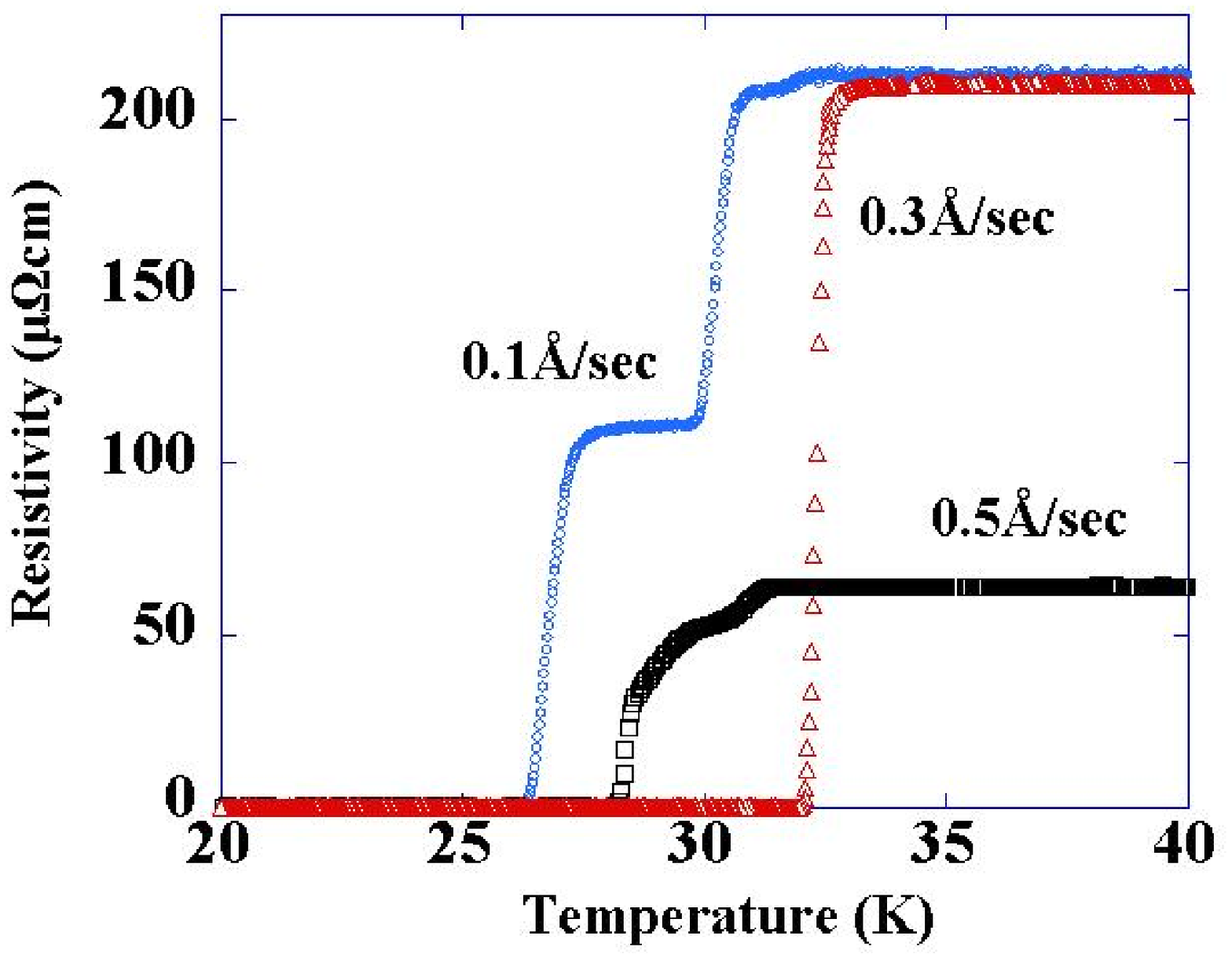}
	\includegraphics[width=0.8\linewidth]{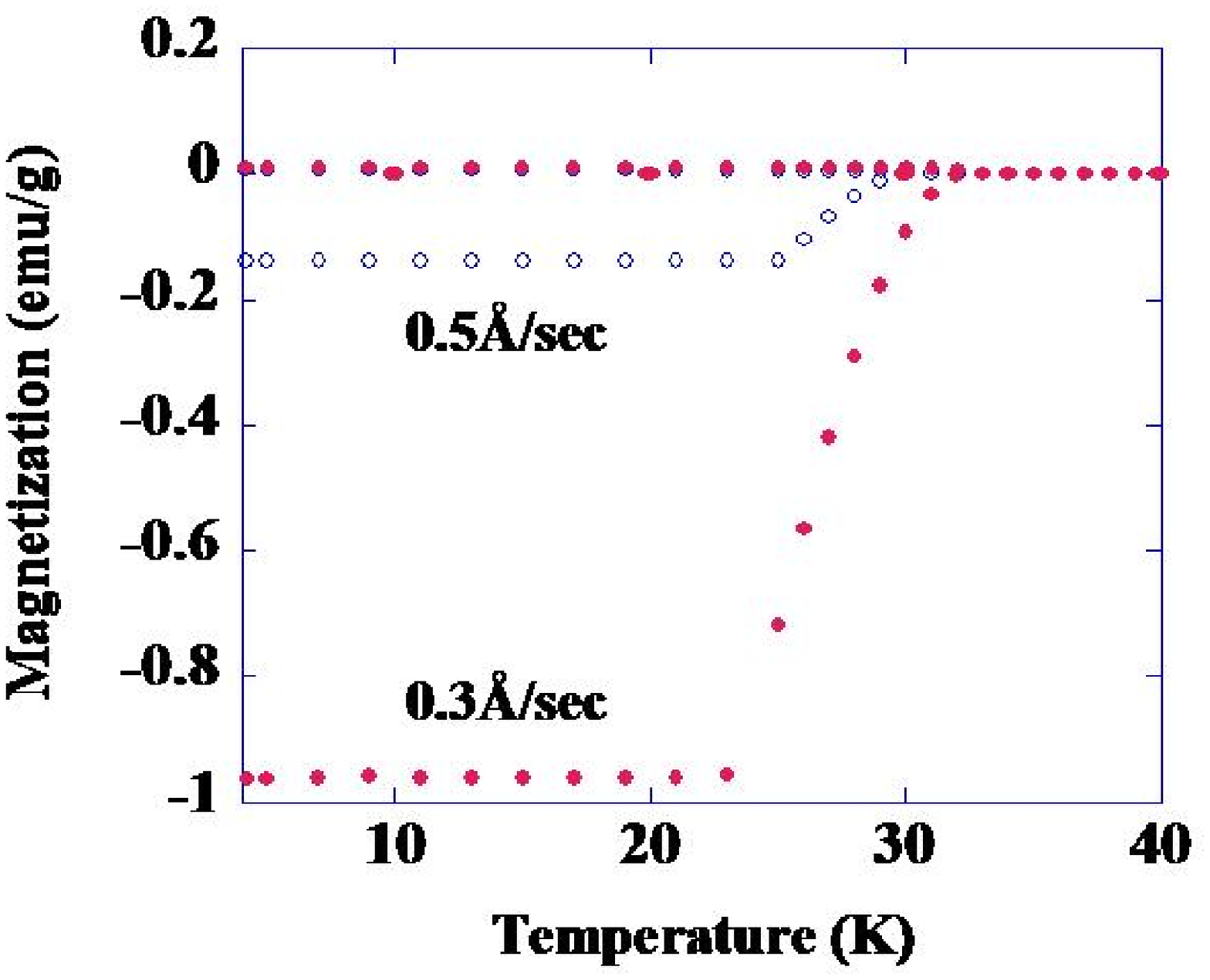}
	\caption{(a) Resistivity as a function of temperature of MgB$_2$ thin films under B deposition rates of 0.1, 0.3 and 0.5\AA/sec. Substrate temperature was fixed at 200$^\circ$C. (b) Magnetization as a function of temperature under B deposition rate 0.3 and 0.5\AA/sec.}
\end{figure}
\begin{figure}[tp]
	\includegraphics[width=0.8\linewidth]{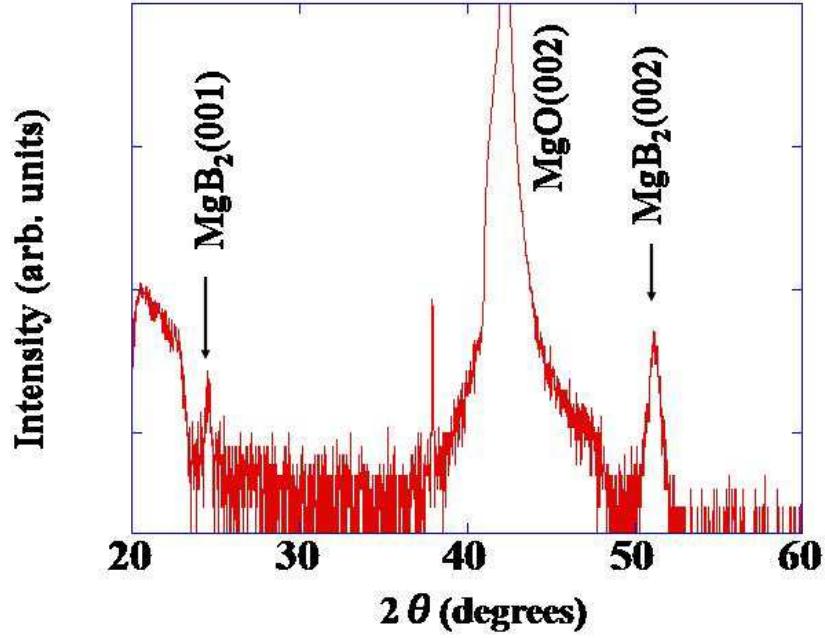}
	\caption{(a) XRD pattern for MgB$_2$ thin films grown on MgO(001) substrate.}
\end{figure}
\begin{figure}[tp]
	\includegraphics[width=0.8\linewidth]{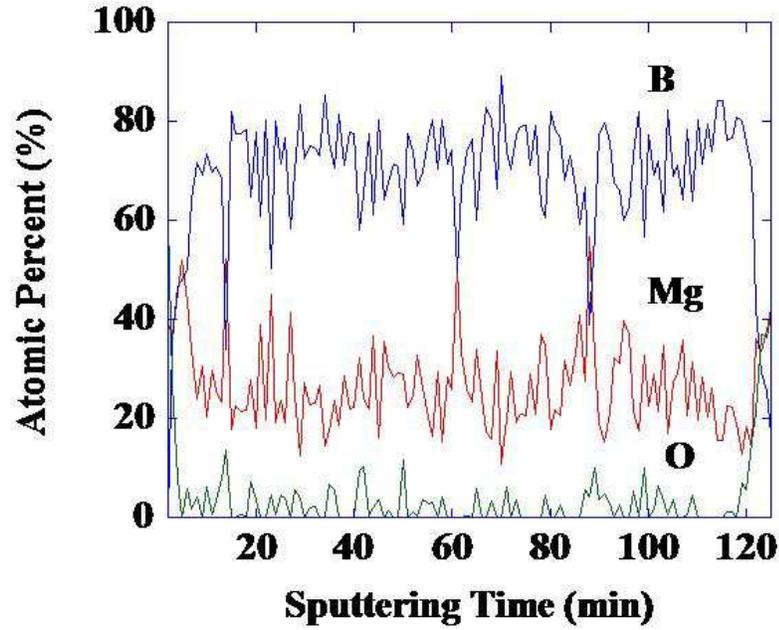}
	\caption{XPS depth profiles of  a MgB$_2$ film, Mg-cap layer was deposited on the surface MgB$_2$ thin films.}
\end{figure}
\begin{figure}[tp]
	\includegraphics[width=0.67\linewidth]{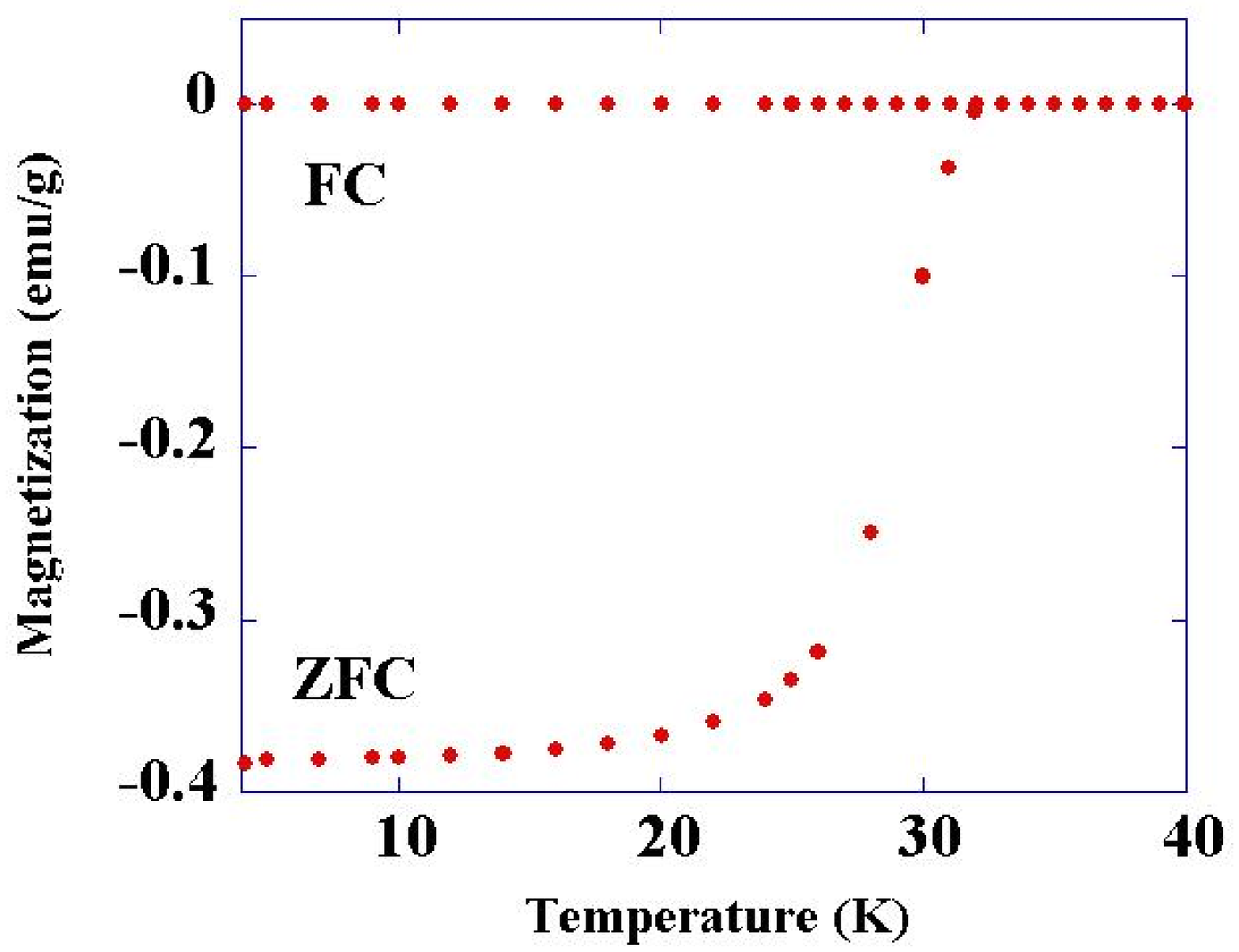}
	\includegraphics[width=0.8\linewidth]{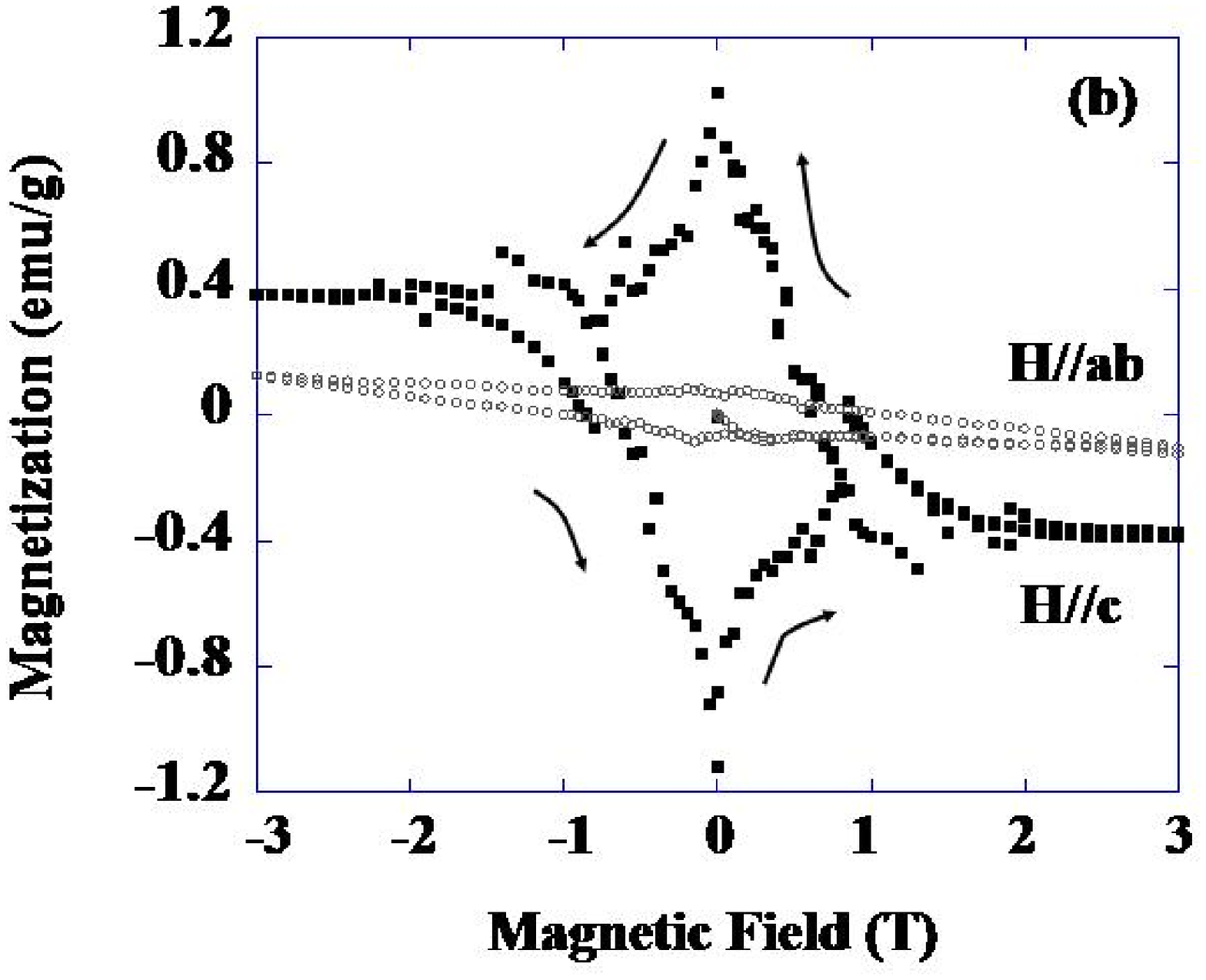}
	\caption{Magnetic properties of the MgB$_2$ thin film. (a) Magnetization as a function of temperature after cooling under zero field and under a field of 1mT at {\it H}$\|$c. (b) Magnetization as a function of applied fields up to 3T at 5K for {\it H}$\|$c and {\it H}$\|$ab.}
\end{figure}
\begin{figure}
	\includegraphics[width=0.8\linewidth]{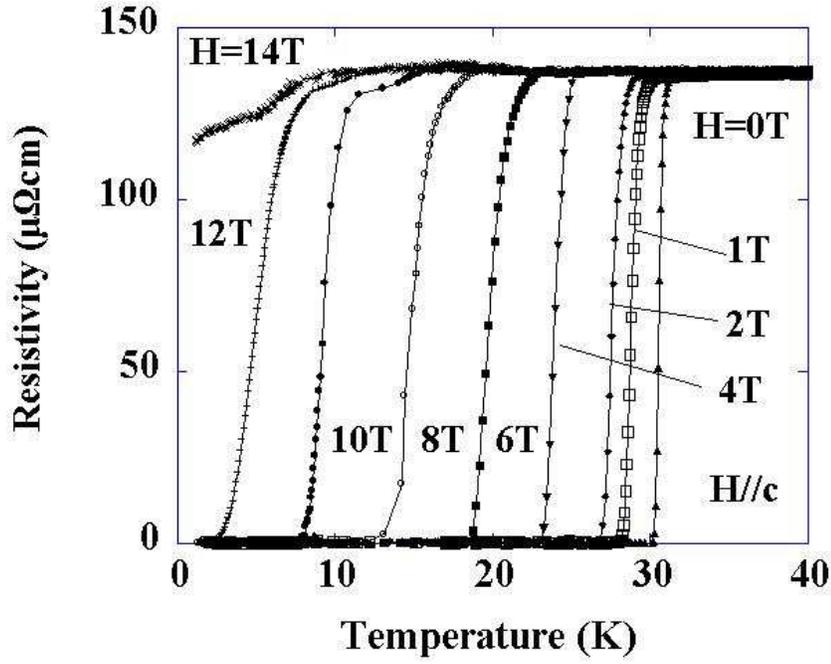}
	\includegraphics[width=0.8\linewidth]{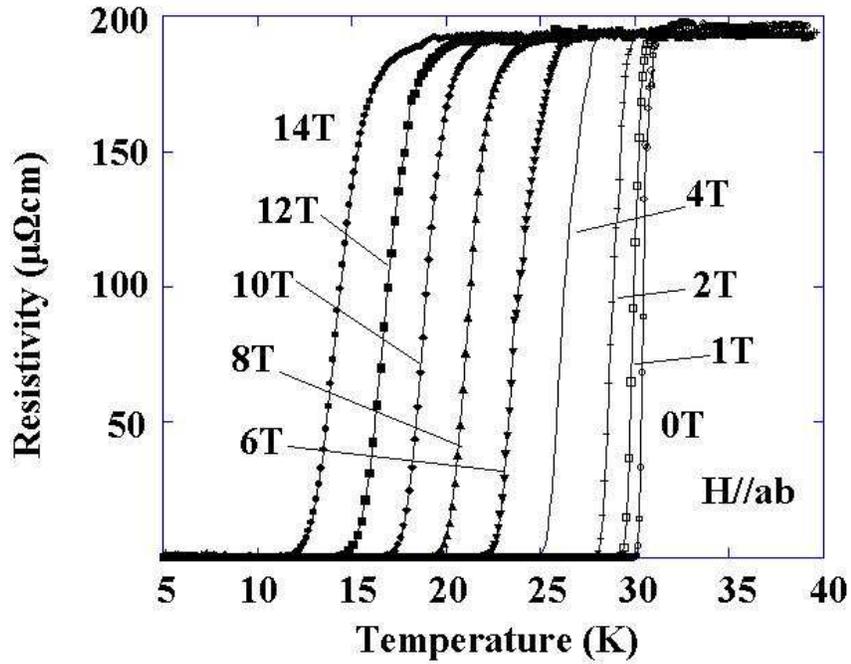}
	\caption{Resistivity as a function of temperature under selected magnetic fields up to 14T for (a) {\it H}$\|$c and  (b) {\it H}$\|$ab.}
\end{figure}
\begin{figure}[tp]
	\includegraphics[width=0.8\linewidth]{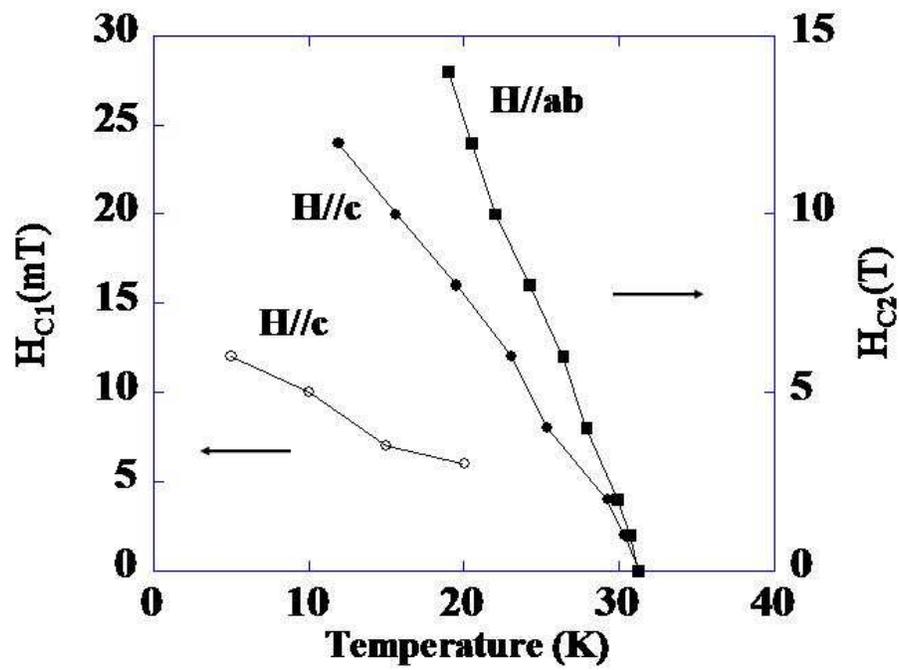}
	\caption{Magnetic field-temperature phase diagram of the as-grown MgB$_2$ thin film determined our transport and magnetization experiments.}
\end{figure}





\end{document}